\documentclass{article}

\usepackage{PRIMEarxiv}

\usepackage[utf8]{inputenc} 
\usepackage[T1]{fontenc}    
\usepackage{hyperref}       
\usepackage{url}            
\usepackage{booktabs}       
\usepackage{amsfonts}       
\usepackage{nicefrac}       
\usepackage{microtype}      
\usepackage{lipsum}
\usepackage{fancyhdr}       
\usepackage{graphicx}       
\graphicspath{{media/}}     
\usepackage{subcaption}
\usepackage{float}
\usepackage{amsmath}

\title{Point Projection Mapping System for Tracking, Registering, Labeling and Validating Optical Tissue Measurements
}

\author{
 Lianne Feenstra, Stefan D.van der Stel, Marcos Da Silva Guimaraes, Theo J.M Ruers and Behdad Dashtbozorg \\
  Image-Guided Surgery \\
  Netherlands Cancer Institute \\
  Amsterdam\\
  \texttt{\{l.feenstra, s.vd.stel, m.d.silva, t.ruers, b.dasht.bozorg\}@nki.nl} \\
}

\begin{document}
\maketitle

\begin{abstract}
Validation of newly developed optical tissue sensing techniques for tumor detection during cancer surgery requires an accurate correlation with histological results. Additionally, such accurate correlation facilitates precise data labeling for developing high-performance machine-learning tissue classification models. In this paper, a newly developed Point Projection Mapping system will be introduced, which allows non-destructive tracking of the measurement locations on tissue specimens. Additionally, a framework for accurate registration, validation, and labeling with histopathology results is proposed and validated on a case study. The proposed framework provides a more robust and accurate method for tracking and validation of optical tissue sensing techniques, which saves time and resources compared to conventional techniques available.
\end{abstract}

\section{Introduction}
\label{sec:Introduction}
Surgery combined with (neo)adjuvant therapy is currently the most common treatment for patients with cancer. Oncological surgery is characterized by a delicate balance between radical tumor resection and sparing as much as possible healthy tissue. For a surgeon, recognizing tumor margins can be challenging since resection of the tumor is mostly based on visual and tactile feedback. This can result in resections too close to the tumor (positive resection margins) or resections too wide from the tumor, leading to increased risk of tumor recurrence, undesired cosmetic outcomes or potential damage to vital anatomical structures. Tumor-positive resection margins vary from 4.3\% in uterine cancer to 35\% in ovarian cancers, and up to 19\% in advanced rectal cancer \cite{Kusters2010PatternsTrial} and 21\% for prostate cancer \cite{Orosco2018PositiveCancers}. In this case, additional treatment such as chemotherapy, radiotherapy, or surgical re-excision may be necessary which affects morbidity as well as the quality of life for patients \cite{Hau2013TheTrial}. In contrast, in breast cancer the excised tissue volume of the resection specimen often exceeds 2-3 times the volume of the tumor, leading to worse cosmetic results \cite{Valejo2013VolumeTherapy,Krekel2011ExcessiveStudy}. Therefore, there is a need for more precise oncological surgery, making it possible to detect tumor regions intraoperatively and thereby lower the number of positive resection margins and additional treatments. 

Optical technologies have shown great potential for the assessment of resection margins since they can reflect the biochemical and functional properties of the measured tissue. These technologies already have been successfully evaluated in multiple oncological domains for discriminating tumor from healthy tissue with high accuracies \cite{deBoer2015Fat/waterBoundaries,Baltussen2020,Langhout2018,Evers2012OpticalTherapy}. This includes point-based measurement techniques such as Diffuse Reflectance spectroscopy (DRS) \cite{deBoer2018TowardsSurgery,BrouwerDeKoning2018}, Raman spectroscopy \cite{Haka2005}, Fluorescence Lifetime Imaging (FLIm) \cite{Alfonso-Garcia2021MesoscopicAccess}, and infrared spectroscopy \cite{Gurjarpadhye2015InfraredDermatology.} as well as image-based techniques including hyperspectral imaging \cite{Kho2021FeasibilitySpecimen, Jong2022DiscriminatingImaging}. Optical tissue sensing technologies have clinical advantages since they are non-destructive, and they do not require exogenous contrast with dyes. Besides, they and have the potential to be performed in real-time, providing immediate feedback to the user.

\begin{figure}[!h]
	\includegraphics[width=0.75\textwidth, height =6.5cm]{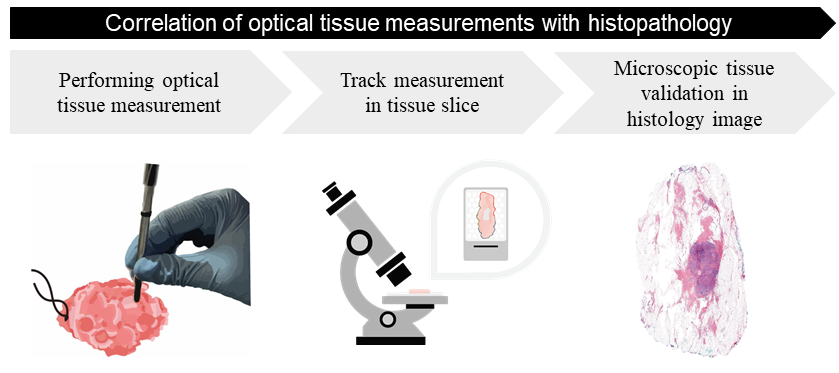}
	\centering
	\caption{Ground truth validation of point-based optical measurements: after the performed optical measurement, a tracking method is needed to trace the performed measurement area back in a gross-sectioned tissue slice. The gross-sectioned tissue slice will be further processed and result in a histology image (H\&E tissue section). From this image, the optically measured tissue area can be defined microscopically and will be considered as ground truth.}
	\label{fig:correlation histopathology}
\end{figure}

The first steps after the development of an optical tool involve ex vivo tissue specimen studies, where the technology will be evaluated for clinical purposes. In order to use optical technologies as a diagnostic tool for the optimization of surgical outcomes eventually, it is important that the optical tissue measurements are validated with a ground truth first \cite{Wilson2018ChallengesImaging}. Ground truth validation of optical tissue sensing technologies is currently provided by hematoxylin and eosin (H\&E) stained tissue sections from which the measured tissue structures can be identified microscopically \cite{Wells2007ValidationView}. From this H\&E section, a pathologist annotates all different tissue structures located in the measured tissue area, which will then be considered as ground truth. Accordingly, it is required to track where exactly on the excised tissue specimen the point-based optical tissue measurements were performed in order to correlate those measurement locations in the gross-sectioned tissue slices and corresponding H\&E-section annotations (Figure \ref{fig:correlation histopathology}). Accurate correlation of an optical tissue measurement to histopathology is especially of importance for the development of (real-time) tissue classification algorithms since incorrect labeling of data will influence the performance during the training of machine learning models. This correlation involves, for example, a registration between a microscopic histology image and a corresponding snapshot image of a tissue specimen.

For the development of accurate tissue classification algorithms and validation of optical tissue sensing techniques, the first step consists of tracking the performed point-based optical measurements on tissue specimens is an important first step. 
Moreover, it has been observed that some studies do not have an adequate tracking method, or they rely on visual correspondence only \cite{Keller2018DiffuseDiodes,Sircan-Kucuksayan2015DifferentiatingDiameters,Skyrman2022DiffuseStudy,Nogueira2021EvaluationDetection}. As a result, correlation with histopathology is based on visual memory and therefore prone to human error. Other studies show conventional approaches to track the position of the optical tissue measurement which involve the placement of ink marks or fiducial markers on the tissue specimen's surface after acquisition \cite{Jong2022DiscriminatingImaging, Lay2016DetectingSpecimens}, the use of measurement grids and live-tracking of the optical probe \cite{Horgan2021Image-guidedDelineation,Gkouzionis2022Real-timeSurgery}. However, these methods are limited since the accuracy of tracking can be affected by human errors and placement of such markers can damage the tissue, complicating histopathologic processing and analysis. For these reasons, it would be desirable to have a more precise and generalized method, applicable to the various optical tissue sensing techniques available, which tracks optical tissue measurements in any desired location without damaging or marking the tissue specimens. 

The second step should deal with the challenge of establishing a robust correlation between the tracked optical tissue measurement locations and the corresponding histopathological tissue labels. Establishing an accurate correlation between optical tissue measurements and ground truth is especially of importance when preparing datasets for training supervised machine learning techniques for tissue discrimination. Using accurately labeled data, tissue classification algorithms can be developed to eventually classify tissue structures in real-time. The labeling of optical data often includes a multistep registration method. Where, for example, a microscopic H\&E section, including tissue annotations from a pathologist (ground-truth), is registered to a white light specimen image \cite{Unger2018MethodAssessment, Lu2014HyperspectralAccess, Halicek2018DeformableDemons}. With this registration, each tracked measurement can be labeled with the definite measured tissue type or tissue type percentages. However, due to histopathological processing such as formalin fixation and paraffin embedding processes, the H\&E sections are generally deformed compared to the optically measured tissue. These deformations include shrinkage, stretching, and compression of the microscopic tissue slices. Sometimes tears and even loss of tissue can be observed as a result of the slicing and staining process. So, simply overlaying images or using affine registration methods between the specimen images and microscopic H\&E sections will therefore be imprecise. Previous studies showed the importance of accounting for tissue deformations when correlating optical tissue measurements with histological results \cite{DeBoer2019MethodDeformations}. Thus, when taking tissue deformations into account, an improvement in the correlation of optical tissue measurements could be achieved. 

In this work, a new framework for accurate validation of point-based optical tissue measurements will be introduced. The first part of this article focuses on the development of a Point Projection Mapping (PPM) pipeline in which we used a custom-built setup and also an off-the-shelf device. With each of these systems, it becomes possible to track and project any number of desirable measurement locations on the tissue specimen without damaging or marking the tissue and works both with optical measurements performed on the surface of tissue specimens as well gross-sectioned tissue slices. Consequently, a generalized method for tracking and registering point-based optical tissue measurements to histopathology will be proposed. With an increased number of accurately labeled measurement locations, time and resources can be decreased since a decreased number of specimens will be required to develop classification algorithms. Also, more accurate data for machine learning will result in better and more robust algorithms. This approach is applicable to multiple specimen types and point-based optical tissue sensing techniques available.

The novel contributions of this paper can be summarized as follows:
\begin{itemize}
    \item Developing a Point Projection Mapping (PPM) system which allows for tracking of point-based optical measurements performed on tissue specimens. 
    \item Introducing a new developed framework for registration, validation and labeling of optical data with histopathology. 
    \item Validating the proposed framework on a use-case scenario. Namely, point-based optical tissue measurements performed on breast cancer lumpectomy specimens.

\end{itemize}
The remainder of this paper is organized as follows: Section \ref{ppm} describes the development and technical information regarding the PPM setups. The proposed framework for the validation of optical tissue sensing technologies will be presented in Section \ref{method}. The results of a use-case scenario are presented in Section \ref{results}, which is followed by the discussion and conclusion in Sections \ref{discussion} and \ref{conclusion}, respectively.

\section{Material and Methods}
In this section, the developed Point Projection Mapping (PPM) setup is introduced first. Afterward, the proposed framework for accurate correlation between optical tissue measurements with histopathology results will be described, by using the PPM setup in a used-case study.

\subsection{Point Projection Mapping}
\label{ppm}
For this study, a PPM pipeline was developed which allows the tracking of point-based optical measurement locations. With such a system, it is possible to project any number of desirable measurement locations on the tissue specimen without damaging or marking the tissue. Optical tissue measurements can be performed on each point projection separately and later be traced back in histology images.

\subsubsection*{Hardware}
We employed two different setups for the PPM system: 1) a custom-built system and 2) the all-in-one HP Sprout Pro.

\begin{figure}[!h]
    \centering
    \includegraphics[width=0.75\textwidth]{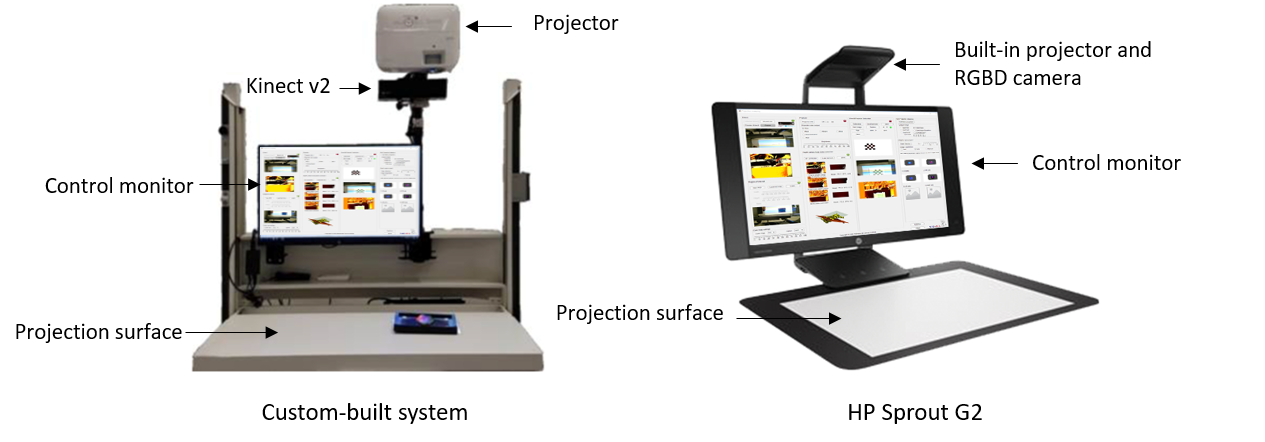}
    \caption{System illustration of custom-built PPMs system on the left and the HP Sprout Pro G2 multimedia system on the right.}
    \label{fig:PPMdevices}
\end{figure}

\textit{Custom-Built setup}\\
Figure~\ref{fig:PPMdevices} illustrates our custom-built setup comprising a standard PC, an RGB-D sensor, and a single projector. The PC is equipped with an Intel(R) Xeon(R) CPU E3-1245 v5 @ 3.50GHz   3.50 GHz, 16 GB of RAM, and an NVIDIA Quadro K620 graphics card. Our choice for the RGB-D camera was the Microsoft Kinect v2, with an RGB camera with a resolution of 1920$\times$1080 pixels, and an infrared camera (depth camera) with a resolution of 512$\times$424 pixels. To facilitate projection mapping, we used a BenQ TH671ST projector with a resolution of 1920 $\times$ 1080 pixels for demonstration purposes. The projector and the Kinect were fixed to an arm facing downward with a distance of 100mm from the surface of the interset.

\textit{HP Sprout}\\
For the PPM system, we also used a HP Sprout Pro G2 multimedia device \cite{2023HP1MU73UAABA}. This device consists of a built-in PC (Intel Core i7-7700T, 16 GB DDR4 Memory, NVIDIA GeForce GTX 960M), a high-resolution DLP projector (1920 x 1280), HP high-resolution downward-facing camera (4416 x 3312), a downward facing RGB-D camera (Orbbec Astra S Mini, RGB Image Resolution: 640 x 480 @30fps Depth Image Resolution: 640 x 480 @30fps) and an integrated 23.8” Touch Display \cite{2023SproutSupport}. 

The software for calibration, 3D image reconstruction and interactive projection mapping for both setups is developed in-house. 

\subsubsection{PPM calibration}
\label{ppm calibration}
An interactive PPM system was designed for real-time surface reconstruction and projection mapping. The RGB-D camera in the setups will provide a stream of depth images as well as corresponding top-view RGB images. The depth images were used for 3D surface reconstruction and the RGB frames were captured and shown to the user on the screen for the selection of points of interest (POI). Furthermore, The projector in the setups was used to illuminate the target surface with bright spots corresponding to the POIs selected by the user.  
However, for such a system, a calibration step is essential for accurate projection mapping. During the calibration process, models will be estimated for the correction and transformation of depth images and extracted meshes to projector coordinates.

\begin{figure}[!h]
    \centering
    \includegraphics[width=0.75\textwidth, height =5.5cm]{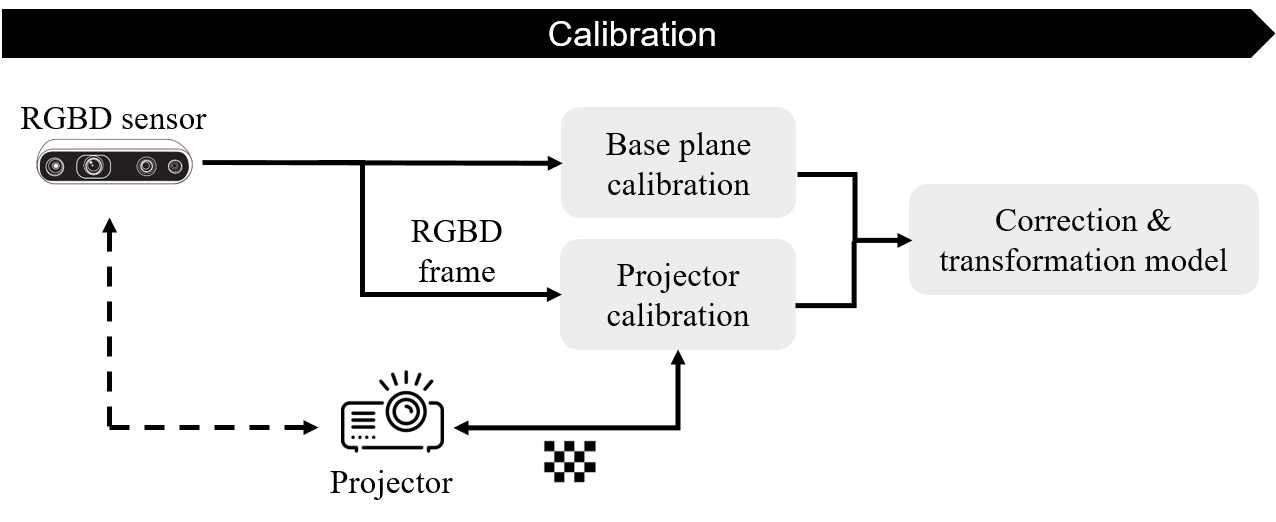}
    \caption{Point projection mapping calibration pipeline.}
    \label{fig:PPMCalib1}
\end{figure}

As demonstrated in Figure \ref{fig:PPMCalib1}, the pipeline of calibration has two phases: 1) Base plane calibration and 2) Projector calibration. It is worth mentioning that the calibration pipeline was identical for both setups.

\subsubsection*{Base-plane calibration}
The built-in RGB-D camera in the HP sprout and Kinect sensor are faced downward and in the case of having a flat surface, the depth camera should return a uniform depth image. However, the captured target surface beneath the camera is not always completely horizontally aligned with the camera's sensor. For an accurate 3D surface reconstruction and projector calibration, a base-plane calibration step is required to discard the deviation caused by an inclined surface.

For the base-plane calibration, a series of depth frames were captured and averaged to reduce any noise presence. Afterward, randomly a set of sample points ($P_i(x_i,y_i,z_i)$) were selected and used to compute the plane that fits best this set of points by calculating the least square of the normal distance to the plane as shown in \ref{eq:ppmcalib1}.

\begin{equation}
\label{eq:ppmcalib1}
    \min \frac{1}{n}\sum_{i=1}^{n} (ax_i+by_i+z_i+c)^2
\end{equation}

Where $a$, $b$ and $c$ are the parameters to minimize the least square error by means of partial derivatives.  
After obtaining the base-plane model, the compensation for deviation of the inclined surface can be done by correcting the depth values for any point ($P_j(x_j,y_j,z_j)$) in a new captured depth frames as shown in \ref{eq:ppmcalib2}.

\begin{equation}
\label{eq:ppmcalib2}
    z_{j}^{new} = z_j +ax_j+by_j+c
\end{equation}

\begin{figure}[!h]
\captionsetup[subfigure]{justification=centering}
        \centering

    \begin{subfigure}{0.49\textwidth}
        \centering
        \includegraphics[width=0.95\textwidth]{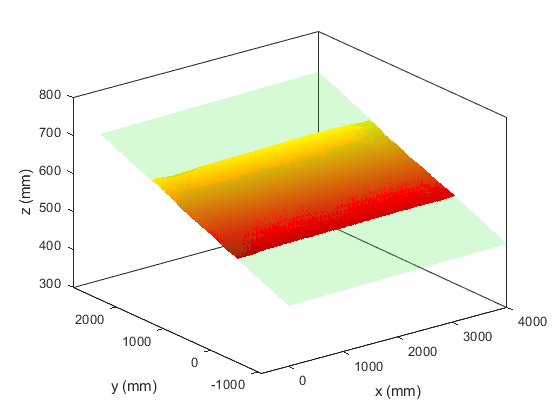}
        \caption{}
        \label{fig:CAM1}
    \end{subfigure}
    \begin{subfigure}{0.49\textwidth}
        \centering
        \includegraphics[width=0.95\textwidth]{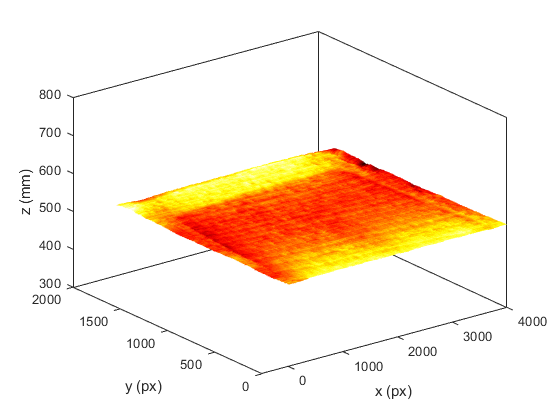}
        \caption{}
        \label{fig:CAM2}
    \end{subfigure}

        \begin{subfigure}{0.49\textwidth}
        \centering
        \includegraphics[width=0.7\textwidth]{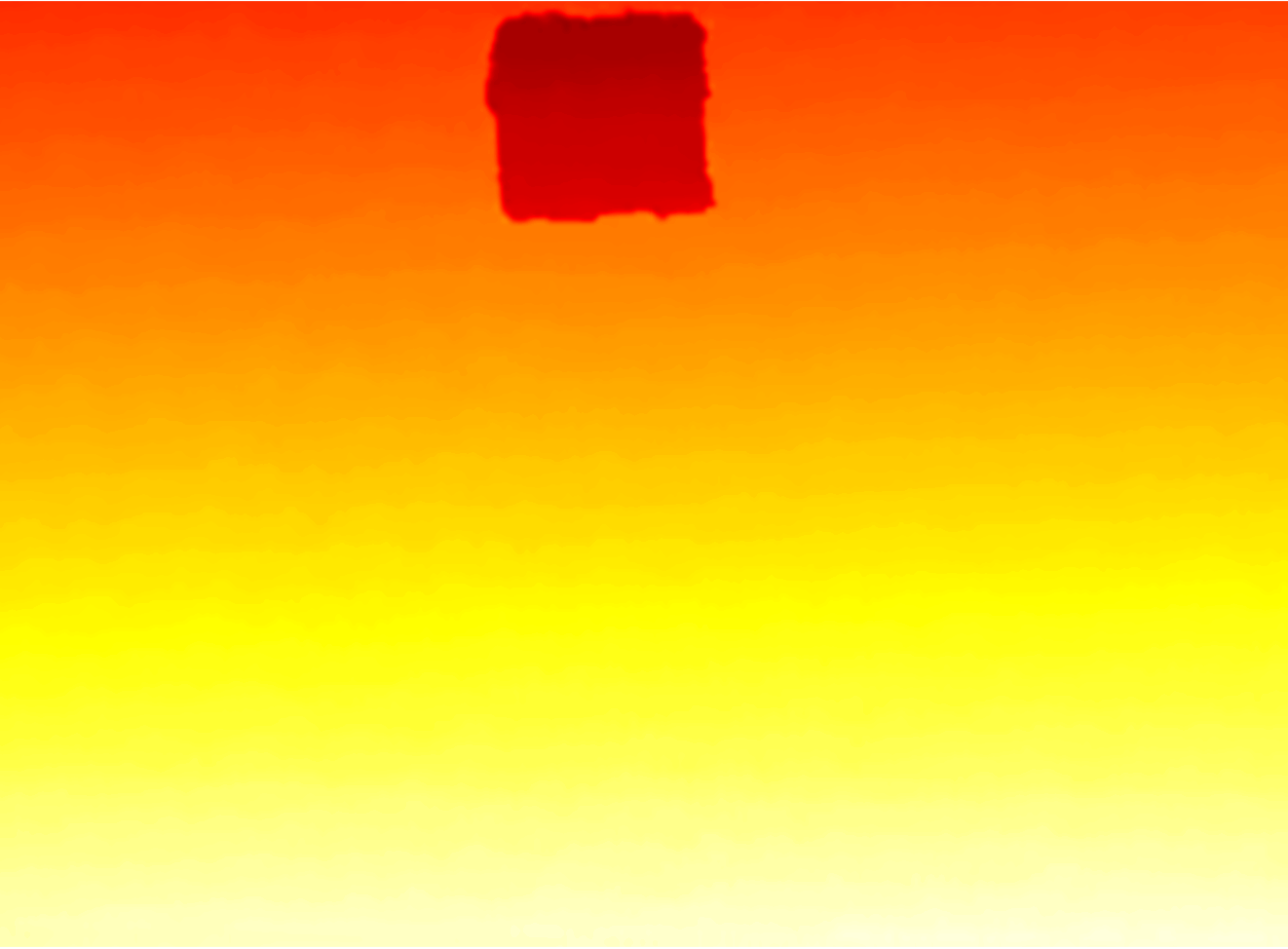}
        \caption{}
        \label{fig:CAM3}
    \end{subfigure}
    \begin{subfigure}{0.49\textwidth}
        \centering
        \includegraphics[width=0.7\textwidth]{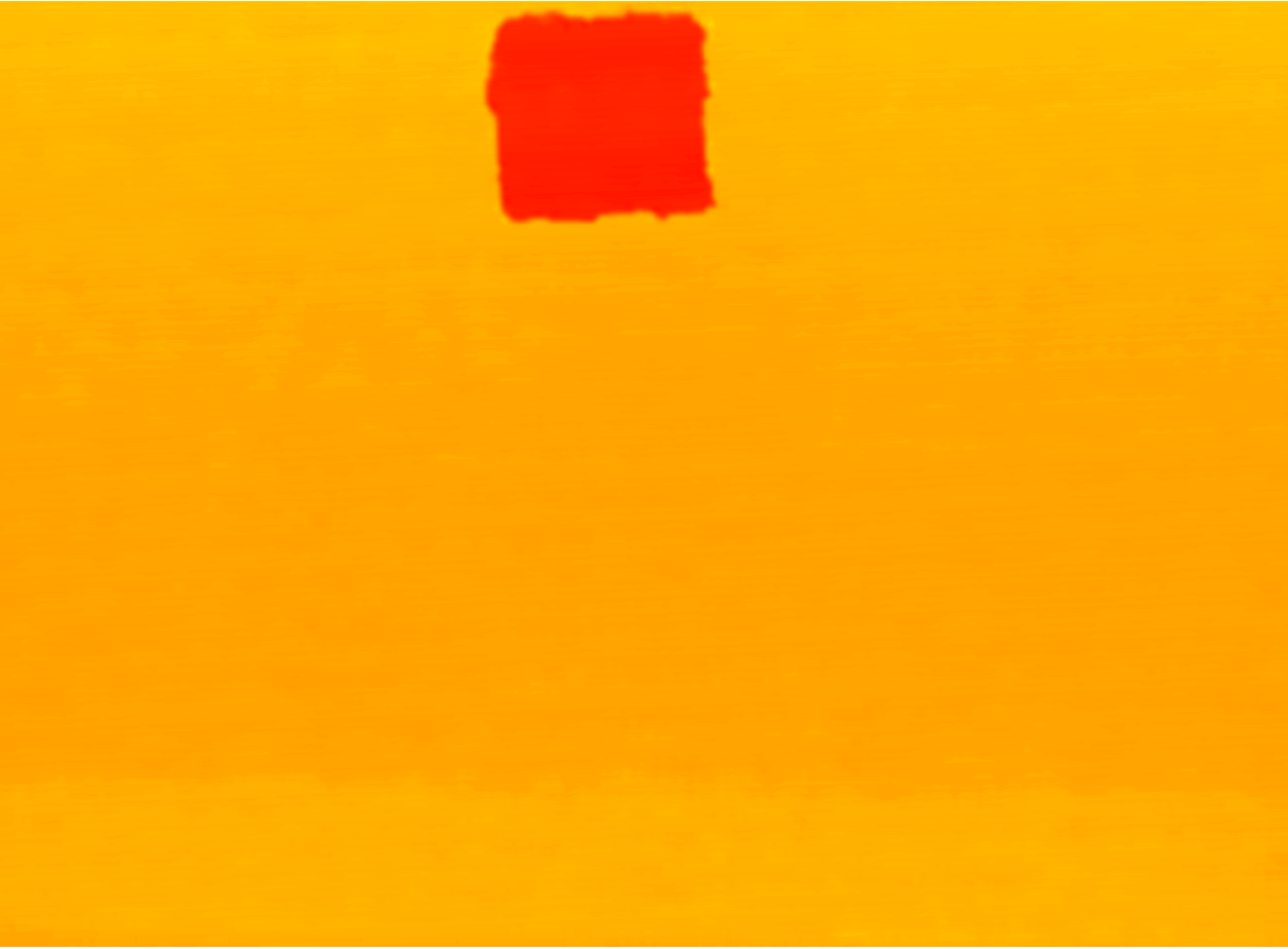}
        \caption{}
        \label{fig:CAM4}
    \end{subfigure}
    \caption{Base-plane calibration: (a) the 3D representation of the camera field of view flat surface before calibration where the green plane represents the plane fitted a set of randomly selected points, (b) 3D representation of the same flat surface after calibration. (c) and (d) an example of a depth frame with an object before and after calibration.}
    \label{fig:PPMCalib2}
\end{figure}

Where $z_{j}^{new}$ is the corrected depth for $P_j$ at spatial coordinate of ($x_j$,$y_j$). The 3D representation of base plane before and after correction as well as an example of captured depth frame with an object is shown in Figure \ref{fig:PPMCalib2}.

\subsubsection*{Projector calibration }
The PPM system requires a precise transformation model to function properly. To address this need, a convenient and efficient calibration approach was deployed that was both fast and easy to execute. 
To implement this approach, two 3D orthogonal spaces with Cartesian coordinate systems were defined: camera space and projector space. In camera space, an arbitrary point is denoted by $P_c(x_c, y_c, z_c)$, while in projector space, an arbitrary point is denoted by $P_p(x_p, y_p, z_p)$. The transformation matrix to convert points between these two spaces is crucial to the projection mapping process, as shown in (\ref{eq:trans}).

\begin{equation} 
\label{eq:trans}
\begin{pmatrix}
\textbf{ R} & \textbf{T} \\ 
 0 & 1 
\end{pmatrix}
\begin{pmatrix}
x_c\\ 
y_c\\ 
z_c\\ 
1
\end{pmatrix}
=
\begin{pmatrix}
x_p\\ 
y_p\\ 
z_p\\ 
1
\end{pmatrix}
\end{equation}

Where $\textbf{R}$ and $\textbf{T}$ denote rotation and translation matrices, respectively. To collect representative sample point pairs in both depth image and screen space for computing the transformation matrix, a 4$\times$5 chessboard pattern (Figure \ref{fig:CAM1}) was utilized and projected onto planes of different heights above the target surface. 
To recognize the sample points in screen space, the sequences of the chessboard pattern were used at various heights and orientations and images were captured by the RGB-D sensor (Figures \ref{fig:CAM2} and \ref{fig:CAM3}). The recognized corner points on the chessboard were then mapped to the depth image by the registration of RGB to the depth images. MATLAB was used to perform the recognition extraction of 12 point pairs per checkerboard configuration to estimate the transformation model. The transformation model was estimated by solving the estimation of the parameters using a derivative-free nonlinear solver. 

\begin{figure}[h!]
\captionsetup[subfigure]{justification=centering}
        \centering

    \begin{subfigure}{0.30\textwidth}
        \centering
        \includegraphics[width=0.95\textwidth]{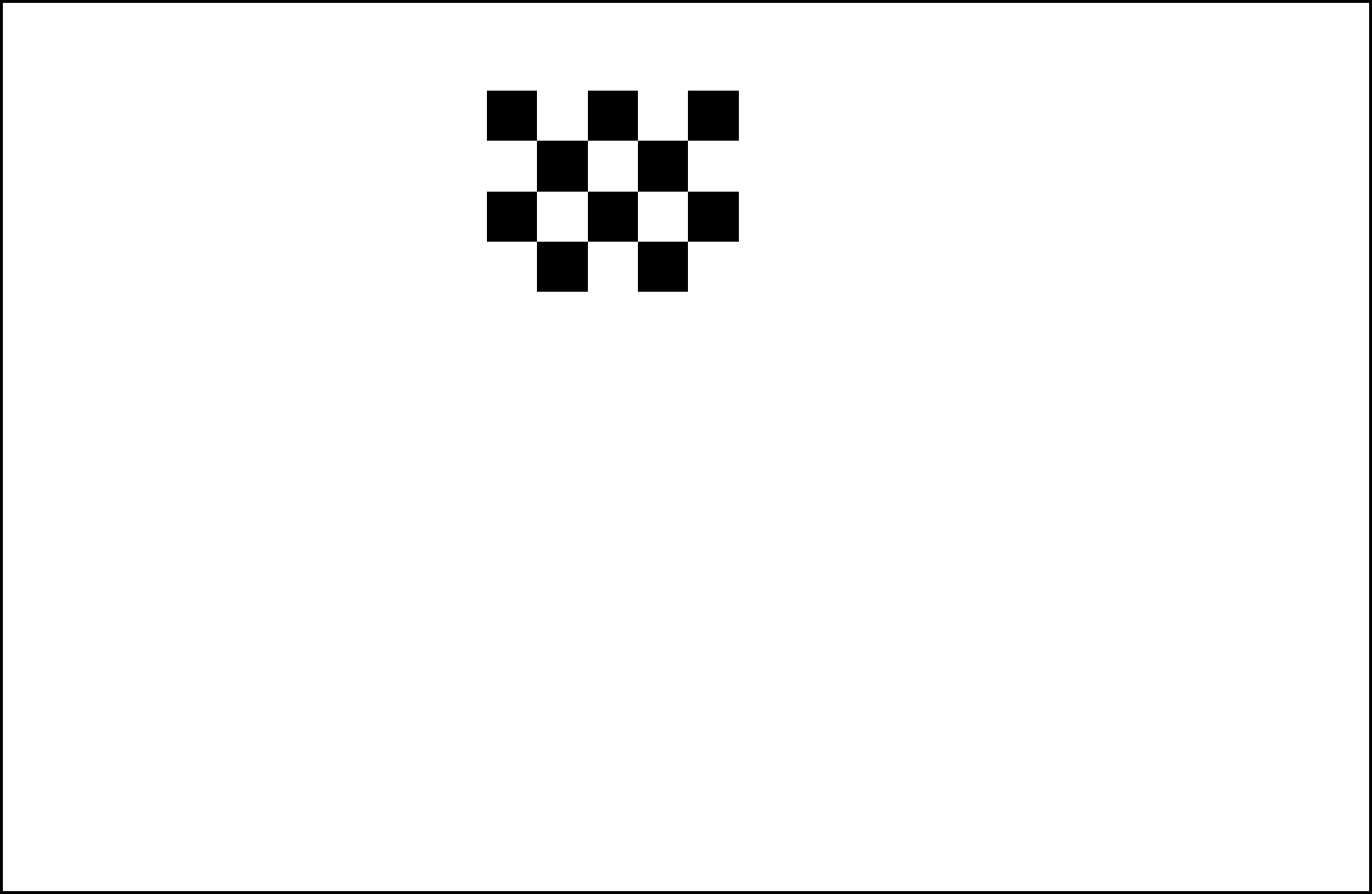}
        \caption{}
        \label{fig:CAM1}
    \end{subfigure}
    \begin{subfigure}{0.30\textwidth}
        \centering
        \includegraphics[width=0.95\textwidth]{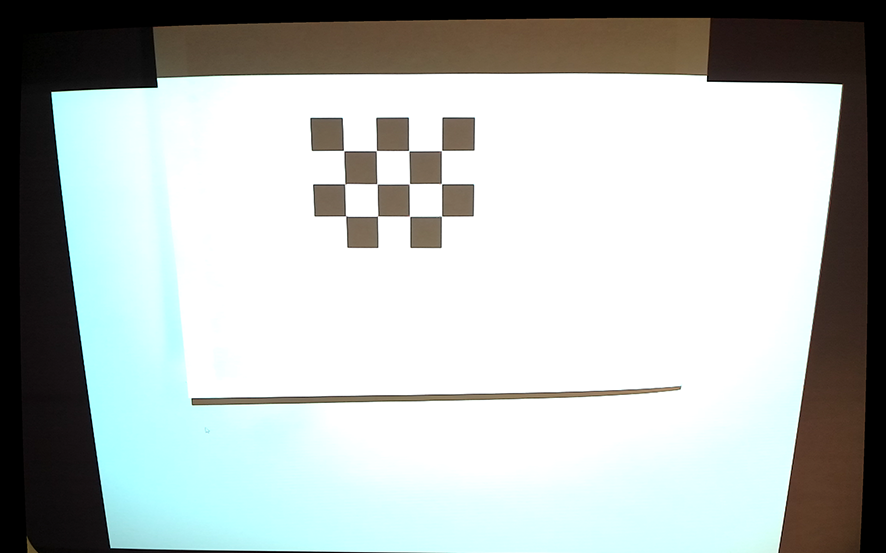}
        \caption{}
        \label{fig:CAM2}
    \end{subfigure}
        \begin{subfigure}{0.30\textwidth}
        \centering
        \includegraphics[width=0.95\textwidth]{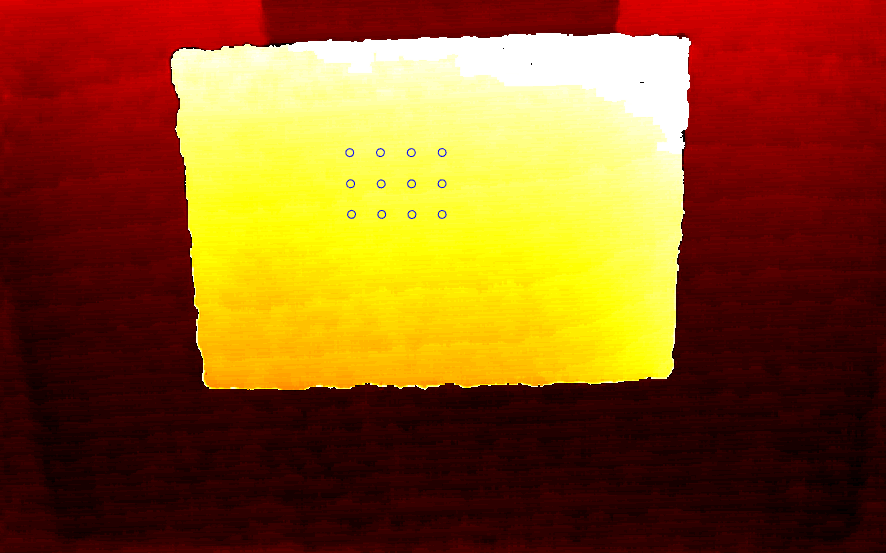}
        \caption{}
        \label{fig:CAM3}
    \end{subfigure}
    \caption{Projector calibration: (a) Checkerboard pattern example, (b) Corresponding acquired RGB image, and (c) Depth image after the projection of the checkerboard pattern.}
    \label{fig:PPMCalib3}
\end{figure}

\subsection{Framework for the validation of optical tissue sensing technologies}
\label{method}
In this section, the developed PPM system will be implemented in a newly introduced framework for registering, labeling and validating optical point-based measurements with histopathology. This framework was evaluated based on a performed use case study. For this, 30 patients who underwent breast-conserving surgery at the Netherlands Cancer Institute – Antoni van Leeuwenhoek (NKI-AVL) were included, and optical point-based tissue measurements were performed on excised lumpectomy specimens. In this specific use case, Diffuse Reflectance Spectroscopy (DRS) measurements were performed using an optical probe. However, this framework can be applied using any other optical point-based technique available. This study was approved by the Institutional Review Board of NKI-AVL and registered under number IRBm20-006 which did not interfere with the standard histopathology processing and subsequent diagnostic procedures. 

\subsubsection{Measurement pipeline}
\label{pipeline}
Figure~\ref{fig:complete pipeline} demonstrates the overview measurement pipeline with three main steps: 1) Specimen collection 2) Selecting, tracking and performing optical measurements and 3) Histology processing.

\subsubsection*{Specimen collection}
\label{subsec:Specimen collection}
Immediately after the performed breast-conserving surgery, the excised lumpectomy specimen was collected in the operating theater from the NKI-AVL hospital and transported to the Department of Pathology. The specimen was inked and gross-sectioned in approximately 5 mm thick tissue slices according to standard protocol until either the tumor area, or the placed Iodine-125 seed became visible (Figure \ref{fig:complete pipeline} a-c). The unsliced part of this lumpectomy specimen was then used for optical tissue measurements. Optical tissue measurements in this study were performed on the inside of the lumpectomy specimens, since the macroscopic appearance of tumor tissue increases the likelihood of performing measurements on tumor sites compared to optical tissue measurements which are performed on the outside of a specimen surface. 

\subsubsection*{Selecting, tracking and performing optical tissue measurements}
\begin{figure}[!h]
	\includegraphics[width=0.75\textwidth]{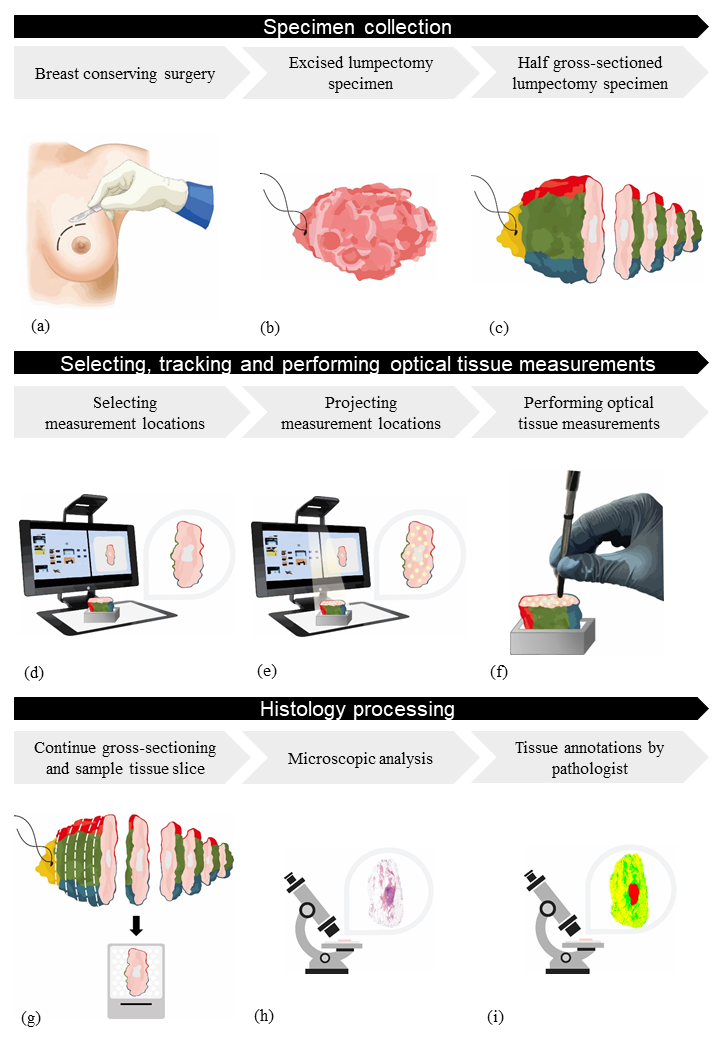}
	\centering
	\caption{Overview measurement pipeline (a) Breast-conserving surgery (b) Excised lumpectomy lump (c) Gross-sectioning of lumpectomy lump until tumor area becomes visible (d) Acquiring snapshot specimen image ($S_O$) with PPM-system and selection of measurement locations (e) Projecting measurement locations and acquiring a snapshot specimen including projected POIs ($S_{POI}$) (f) Performing DRS measurements (g) Continue gross-sectioning and sample measured tissue slice (h) Processing and acquiring histology image ($H_O$) (i) Ground truth tissue annotations by pathologist and acquiring annotated histology image ($H_A$).} 
	\label{fig:complete pipeline}
\end{figure}

The half-sliced lumpectomy specimen was positioned in a fixed holder and placed in the field of view of the PPM system. A macroscopic top-view snapshot image of the specimen was acquired and displayed on the screen (Figure \ref{fig:complete pipeline} d). From this image, points of interest (POIs) were selected manually. The number of points can be adjusted depending on the size of the specimen. After selection, the POIs were projected as light dots on the specimen's surface. Next, a new macroscopic top-view snapshot image of the specimen, including projected POIs, was acquired by the PPM system (Figure \ref{fig:complete pipeline} e). The diameter of the projected POI can be adjusted to the size of the used optical probe. After these series of steps, the PPM system outputs two different specimen images: a snapshot specimen image ($S_O$) and a snapshot specimen image including projected POIs ($S_{POI}$). 

After projecting the POIs on the specimen's surface, optical point-based tissue measurements were performed on each predefined location separately (Figure \ref{fig:complete pipeline} f). After positioning the probe on the POI correctly, the projector from the PPM system can be turned off so that the projected light is not interfering while performing optical tissue measurements.

\subsubsection*{H\&E processing}
Next, the remaining half-sliced lumpectomy specimen was further processed by the Department of Pathology, where sagittal slicing and gross sectioning of the lumpectomy specimen continued. The measured tissue slice, the surface on which the optical tissue measurements were performed, was then placed in a megacasette (Figure \ref{fig:complete pipeline} g.). According to standard protocol, a microscopic H\&E section was created and digitalized with Aperio® ScanScope AT2 (Leica Biosystems, Wetzlar, Germany) (Figure \ref{fig:complete pipeline} h). All histology images were uploaded to Slide Score (web viewer for high-resolution scans of microscopic histopathology slides). Here, for each microscopic H\&E image, invasive carcinoma, ductal carcinoma in situ (DCIS), connective - and fat tissue were annotated by a pathologist and considered as ground truth (Figure \ref{fig:complete pipeline} i). 
After finalizing the complete histopathology processing of the lumpectomy specimen, two different microscopic images were generated: a histology image of the measured breast specimen ($H_O$) and an annotated histology image of the measured breast specimen ($H_A$). 

\subsubsection{Correlation with histopathology}
To summarize, after completing the measurement pipeline, four different images were obtained: two snapshot specimen images ($S_O$ and $S_{POI}$) and two histology images ($H_O$ and $H_A$). These images will be used in the following registration pipeline to correlate the snapshot specimen image (including POIs) with histopathology. The histology image (including annotations of the pathologist) was used to label each optical tissue measurement with the correct pathology label.

\subsubsection*{Automatic deformable image registration}
\label{registration}
\begin{figure}[!h]
	\includegraphics[width=0.75\textwidth]{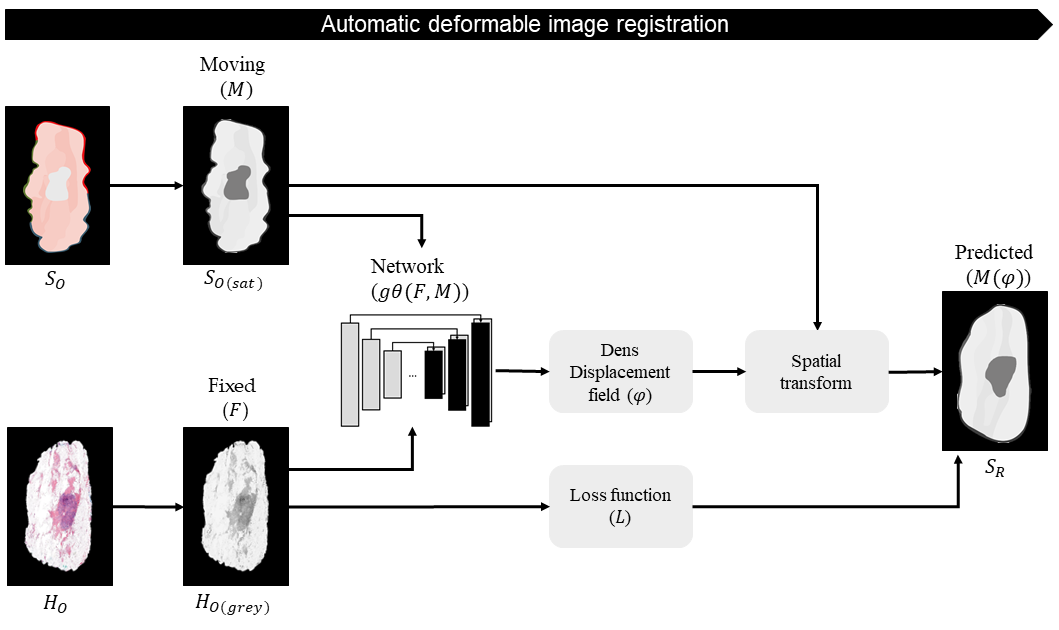}
	\centering
	\caption{Automatic deformable image registration: $H_O$ is converted to single gray scale image ($H_{O(grey)}$). For more similarity of intensity levels $S_O$ is converted to grayscale by using the saturation values only ($S_{O(sat)}$). These images are used as input for the unsupervised deep convolutional neural network ($g_\theta(F, M)$) with fixed histology image $H_{O(grey)}$ (F) and a moving snapshot specimen image $S_{O(sat)}$ (M). Mutual information is used as loss function (L). The network outputs a dense displacement field (DDF($\varphi$)) which defines the mapping from moving image coordinates to the fixed image and is used to register $M$ with $F$. This results in predicted image $S_R (M(\varphi)$).} 
	\label{fig:pre and deformable}
\end{figure}

In a previous study, an unsupervised deep learning-based deformable multi-modal image registration method was developed which is able to account for deformations between images from different modalities [Ref other paper]. The architecture of this automatic deformable image registration method is based on the VoxelMorph principle and uses a deep convolutional neural network ($g_\theta(F, M)$), similar to UNet \cite{Balakrishnan2018AnRegistration,Balakrishnan2019VoxelMorph:Registration}, as displayed in Figure \ref{fig:pre and deformable}. The model uses two input images, in this case, a fixed microscopic histology image (F) and a moving snapshot specimen image (M), which can be switched for own preferences. Since this network was trained with two-channel images input images, it is required to convert $H_O$ and $S_O$ to single grayscale images. To create more comparable intensity levels between both images, the macroscopic top-view specimen image was converted to grayscale by using the saturation values only, as showed in Figure \ref{fig:pre and deformable}. Both input images were resized to 256 by 192 pixels to reduce the computational effort of the network.

The output of the model consists of a dense displacement field (DDF). This DDF has the same size as the moving image and can be defined as a set of vectors that displays the displacement of each individual pixel of this moving image. Thus, the DDF ($\varphi$) defines the mapping from moving image coordinates to the fixed image and was used (in combination with a spatial transform function) to register both images which results in the predicted image (M($\varphi$)). Mutual Information was used as a loss function (L) which is a common objective function for the computation of the similarity between two images acquired in different modalities.

For all 30 lumpectomy specimens, the Dice score and mutual information were calculated between the registered and unregistered images to evaluate the performance of the automatic deformable registration model. Dice score is a commonly used metric in image registration that measures the similarity between two binary images based on the alignment of two images. The Dice score ranges from 0 to 1, where 0 indicates no overlap, and 1 indicates a complete alignment between the reference and registered image. This metric is mostly evaluating the shape of an image. And since a deformable registration is applied, it is also important to evaluate the overlap of the central regions in the images. This can be achieved by calculating the mutual information (MI) between two images. The basic idea of MI in image registration is to measure the similarity between two images by comparing the histograms of these images. The MI between two images is the amount of information that is shared between their histograms. Specifically, it measures how much the joint histogram of the two images deviates from the product of their individual histograms. Thereby, determining the optimal alignment of two images by finding the transformation that maximizes the mutual information between them. A high MI value indicates that the images are similar and easier to align, while a low MI value indicates that the images are dissimilar and more challenging to align. 

Statistical analysis was performed using IBM SPSS statistics v27 (SPSS Inc., United States). Normal distribution was assessed with the Shapiro-Wilk test. Statistical analysis for normally distributed data was performed with an unpaired t-test, and for non-normally distributed data using a Mann-Whitney test. Whereas, a p-value $\leq$0.05 was considered statistically significant. 

\subsubsection*{Label extraction for tissue classification}
In order to extract tissue labels for each measurement location, it is necessary to track the measurement locations in the annotated histology image (ground truth). Therefore, the first step was to extract all measurement locations from $S_{POI}$. X- and Y- coordinates of the centers of these objects were determined and a new binary image with center points was created. Next, the measurement areas were imitated by creating circles corresponding to the size of the used optical probe (which can be adjusted based on the probed volume). Since $S_{POI}$ has the same orientation as the input image $S_O$, the output DDF can be used to apply the obtained deformable registration to the snapshot specimen image including POIs. In this case, the DDF was applied to the binary image, with the same size as $S_{POI}$, to transform the extracted measurement areas to the correct orientation. By overlaying the annotated histology image $H_A$ with the registered binary image (with extracted measurement locations), the optically measured tissue types are visualized for each measurement location microscopically and can be considered as ground truth. The last step, involves the process of creating labels by calculating tissue type percentages for every tracked and registered measurement location. 
In this study, we choose a microscopic histology image as fixed image (F) since it is easier to apply a DDF on measurement locations compared to a microscopic structured when extracting tissue labels. However, this order can be changes to own preferences.


\section{Results}
\label{results}

\subsection{Evaluation PPM-system}
The accuracy of the PPM system is calculated after the calibration procedure. The root mean square error (RMSE) of the transformation model was estimated by the difference between sampling points and the mapping results using a checkerboard. 
The overall system error of the custom-built Kinect-projector setup was ~0.59 mm. For the HP sprout system, this resulted in an RMSE of ~0.15mm. The difference in error can be due to differences in both the depth camera resolution and device stability. In the case of the HP Sprout, the projector and RGB-D camera are integrated and fixed in place, providing greater stability. However, in the custom-built system, while the projector and sensor are also fixed, they are still vulnerable to slight movements, which may impact the calibration which could result in lower accuracy.

\subsection{Acquired images and input images}
\begin{figure}[!h]
	\includegraphics[width=0.55\textwidth]{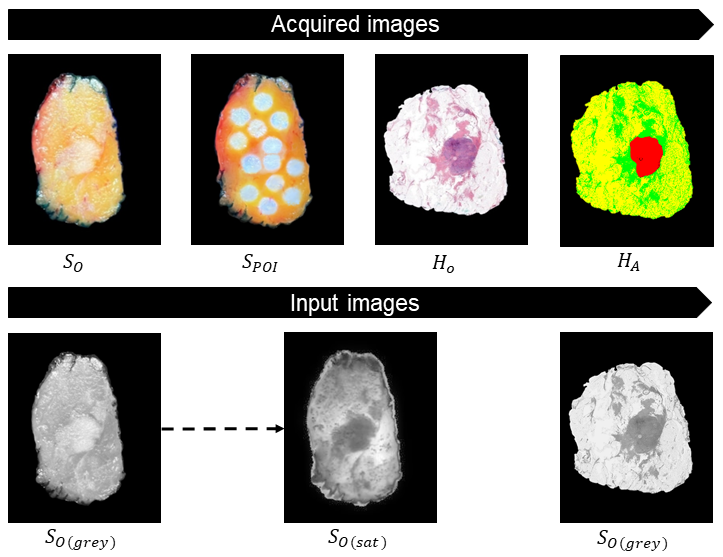}
	\centering
	\caption{Example of the acquired images: the macroscopic top-view snapshot  image of the lumpectomy specimen with and without projected POIs ($S_O$ and $S_{POI}$) and the microscopic histology image with and  without annotations ($H_O$ and $H_A$). Both input images $H_O$ and $S_O$ were converted to single gray scale images ($H_{O(grey)}$ and $S_{O(grey)}$). For more similarity of intensity levels $S_{O(grey)}$ is converted to saturation values only ($S_{O(sat)}$)}
	\label{fig:result registration}
\end{figure}

Optical tissue measurements were obtained from 30 lumpectomy specimens, for which we completed the whole pipeline as described in Section \ref{pipeline}. This resulted in four different images for each specimen:  $S_O$, $S_{POI}$, $H_O$ and $H_A$. Before using the automatic deformable image registration, the input images $S_O$ and the microscopic histology image $H_O$ were converted to grayscale. By using only saturation values, $S_O$ obtained similar intensity levels as $S_O$ (Figure \ref{fig:result registration}).

\subsection{Automatic deformable image registration}
\begin{figure}[!h]
	\includegraphics[width=0.45\textwidth]{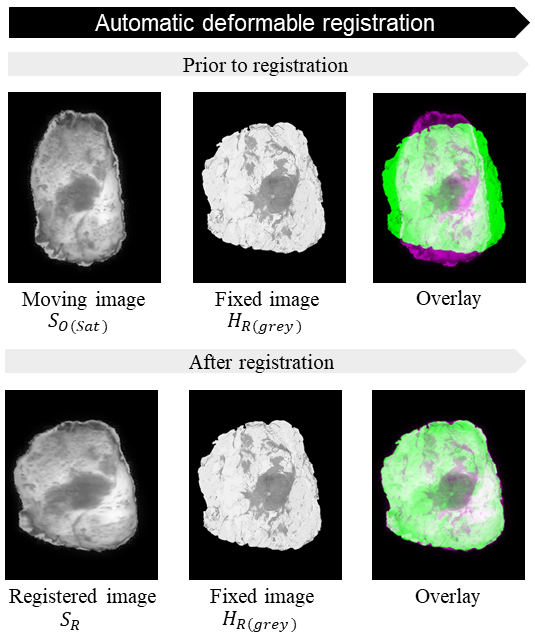}
	\centering
	\caption{Performance automatic deformable registration. Prior to registration: overlay between moving image $S_{O(sat)}$ (purple) and fixed image $H_{O(grey)} (green)$. After registration: overlay between predicted image $S_{R}$ (purple) and fixed image $H_{O(grey)}$ (green)}
	\label{fig:results registration 2}
\end{figure}

\begin{figure}[!h]
\captionsetup[subfigure]{justification=centering}
        \centering
     \begin{subfigure}{0.49\textwidth}
        \centering
        \includegraphics[width=0.9\textwidth]{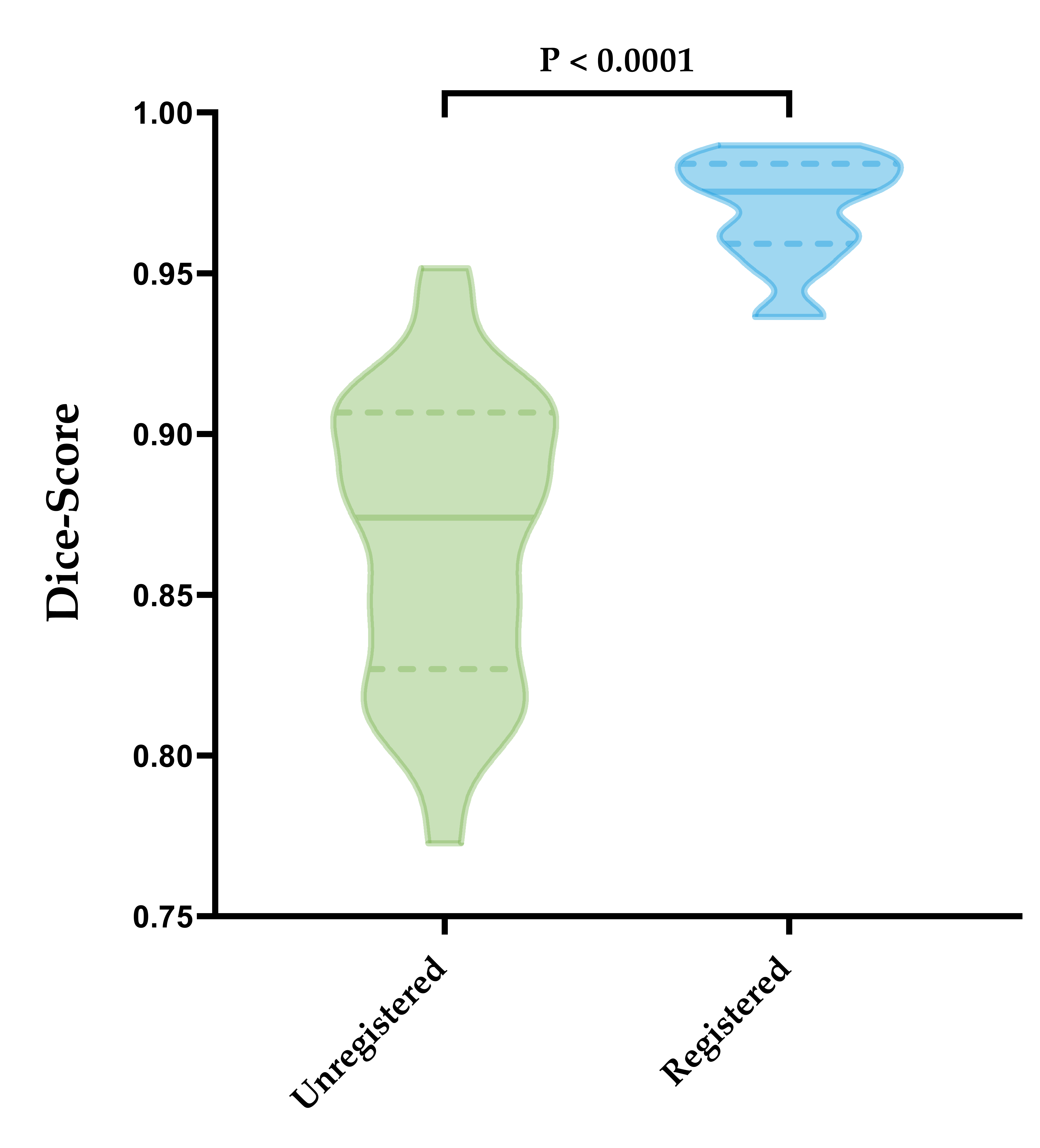}
        \caption{}
    \end{subfigure}
    \begin{subfigure}{0.49\textwidth}
        \centering
        \includegraphics[width=0.9\textwidth]{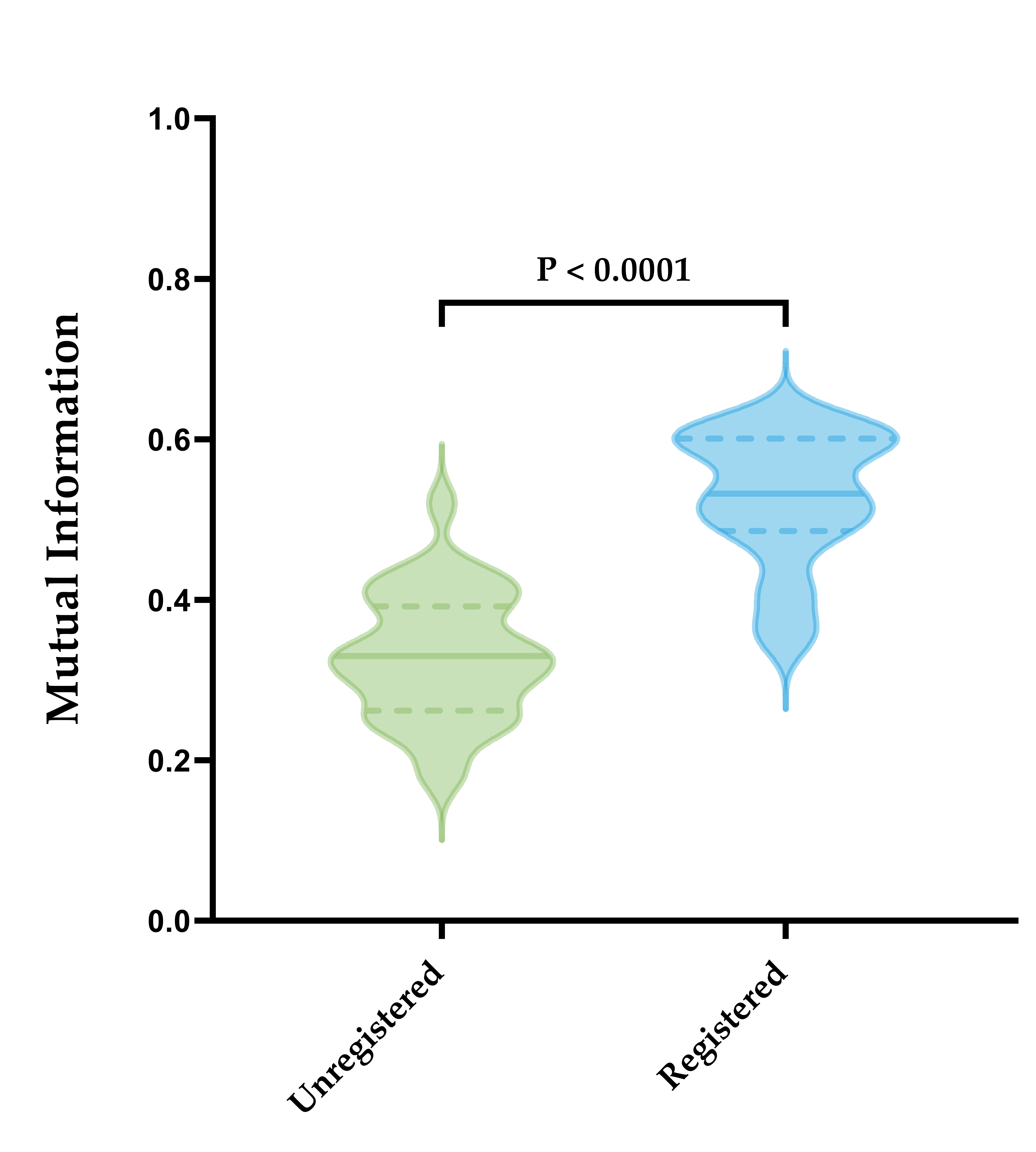}
        \caption{}
    \end{subfigure}
    \caption{Evaluation automatic deformable image registration method (a) Dice score (b) Mutual information. Green and blue visualize the distribution of unregistered and registered dataset, respectively. The middle line represents the median, whereas the thinner dotted lines represent the interquartile range (IQR).}
    \label{result graphs}
\end{figure}

Figure \ref{fig:results registration 2} shows the example for the overlap between the input images, before and after, the automatic deformable image registration was applied. The results for both Dice score and MI are visualized in Figure \ref{result graphs}. 

The violin plots show the distribution for all 30 lumpectomy specimens, before and after the registration was applied. The width of these plots shows the relative frequency in which each value occurs, and becomes wider when the value occurs more frequently and with a higher probability. The distribution for the unregistered Dice score images ranges between 0,77-0,95 (median 0,86 $\pm$ 0,05) and 0,94-0,99 (median 0,97 $\pm$ 0,02) after registration was applied. Whereas, the distribution for the mutual information images ranges between 0,17-0,52 (median 0,33 $\pm$ 0,08) and 0,34-0,63 (median 0,52 $\pm$ 0,08) for the unregistered and registered images respectively. 

\subsection{Label extraction for tissue classification}
\begin{figure}[!h]
	\includegraphics[width=0.75\textwidth, height=6.5 cm]{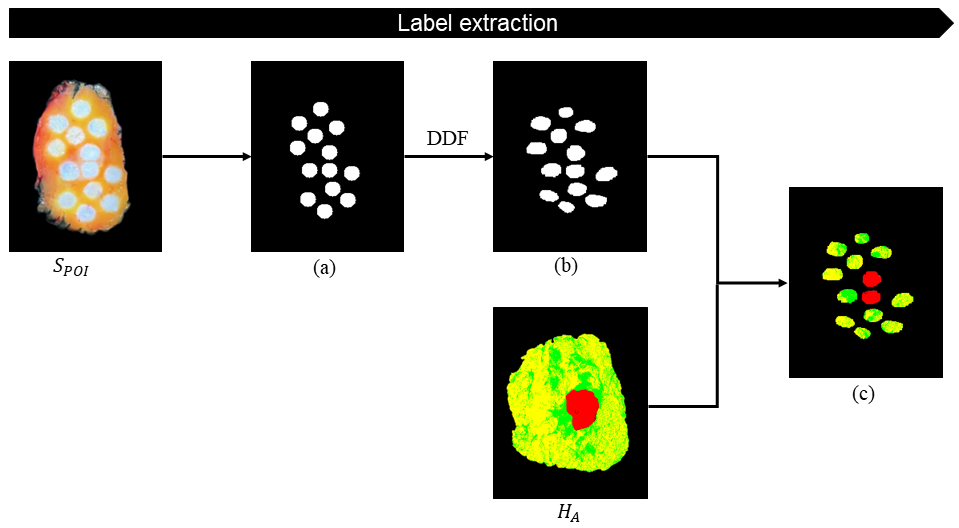}
	\centering
	\caption{Pipeline label extraction (a) Binary image of extracted measurement locations. $S_{POI}$ has the same orientation as input image $S_{O(sat)}$ so the DDF can be applied on the binary image, which results in a registered binary image with extracted measurement areas (b). The annotated histology image ($HE_A$) (where yellow, green and red represents fat, connective tissue and invasive carcinoma respectively) has the same orientation as the input image $H_O$. Therefore, $H_A$ can be overlayed with the registered binary image with extracted measurement areas resulting in tissue label percentages used as ground truth (c)}
	\label{fig:labels}
\end{figure}

The specimen image with projected POIs ($S_{POI}$) has the same orientation as input image $S_{O(sat)}$. Thus, the output DDF can be applied to a binary image with extracted measurement locations to register all locations with histopathology. Therefore, the registered binary image with extracted measurement areas was overlaid with the annotated histology image ($H_A$) in order to determine tissue labels percentages used as ground truth. All steps of the framework for label extraction are visualized in Figure \ref{fig:labels}.

\section{Discussion}
\label{discussion}
Validation of optical tissue sensing techniques is necessary before these technologies can be implemented in diagnostic tools and provide real-time tissue classification during surgical procedures. To make the performance of classification algorithms as accurate as possible, a precise method for tracking the performed optical measurements on tissue specimens is crucial. Such method should enable that measurement areas can be traced back in microscopic tissue sections and may serve as ground truth tissue labels. However, due to histopathology processing, accurate correlation between optical tissue measurements with microscopic tissue sections is often hampered by tissue deformation. In this study, a newly developed framework is introduced for improved tracking, registering and labeling of optical tissue measurements which provides further validation for their clinical applicability. With the use of a Point Projection Mapping (PPM) system, projection of measurement locations on the tissue specimen becomes possible. Acquired top-view specimen images ($S_O$, $S_{POI}$) were used for the following correlation with histopathology. Using an unsupervised automatic deformable multi-modal image registration method, measurement locations can be traced back in the annotated histology images ($H_A$). Labels are created by calculating the percentages of involved tissue types for each tracked and registered measurement location. 

A registration between the tracked optical tissue measurement locations and histopathology is needed to create ground truth tissue labels. Therefore, in this case-study, an automatic deformable registration was applied on a new acquired dataset of optical tissue measurements of 30 lumpectomy specimens to assess the registration performance. The distributions of obtained Dice and MI for the registered images were significantly higher compared to the ones obtained from the unregistered images (Figure \ref{result graphs}). For the Dice, the majority of the images after the registration were distributed with a median of 0,96 $\pm$ 0,01 as visualized in Figure \ref{result graphs} a.  Meaning, based on the general shape of the images, an accurate overlap is achieved. MI was used to quantify the similarity between different image modalities and was calculated using histograms of the images and the joint probability distribution of their intensity values. The majority of unregistered images are distributed around the median of 0,33 $\pm$ 0,07. Whereas, the majority of the cases are located above the median of 0,52 $\pm$ 0,08 after registration meaning an improved alignment of the inside structures was achieved, as visualized in Figure \ref{result graphs} b. MI is originally used for comparing single modality images. However, in this study, we are dealing with registration between different image modalities with different gray intensity distributions. Although MI gives an impression of an improvement in overlaying structures (registration) it is not the most optimal metric to access the registration performance between multi-modal images.

The first step in the validation of technologies for optical tissue sensing involves the tracking of measurement locations. The developed PPM system showed a very high precision when projecting measurement locations on the lumpectomy specimens (RMSE of~0.15mm using HP Sprout device) and thereby demonstrates added value for implementation in the proposed validation framework. 
It is important to note that we also utilized a custom-built device in our experiments, which yielded slightly lower but similar performance (RMSE of ~0.59). This custom-built device can be readily reproduced using any RGB-D camera and projector, addressing the concerns about the limited availability of the HP Sprout Pro G2 multimedia device. The differences in accuracy between these two systems emphasize the importance of factoring in both depth camera resolution and the sturdiness of the integration RGB-D camera and projector when building such a PPM system.

To the best of our knowledge, this is the first automated tracking system using projection mapping that minimizes tracking errors compared to other methods, for example, the use of ink to mark the measurement locations. The accuracy of the ink placement can involve a human error since the locations will be marked after the measurements are performed \cite{Baltussen2019ComparingCancer,Kho2021FeasibilitySpecimen, Jong2022DiscriminatingImaging, Sharma2012Auto-fluorescenceDetection,Lay2016DetectingSpecimens}. Placing ink marks prior to the measurements is not feasible, since the ink can be observed in the spectral data. Besides, placed ink marks can diffuse to the surrounding region, resulting that the mark is not exactly representing the exact measurement location. This issue also limits the number of measurements at possible points of interest, since ink marks with the same color are not distinguishable. When measurements are performed too close together, the ink marks will overlap, which makes it even impossible to track the separate measurement locations back in the corresponding histology image. Besides, this approach is not applicable for optical measurements performed on gross-sectioned tissue slices, since the placed ink will fade out during the following histopathological processes. In this case, the use of permanent fiducial markers (for example, small burn marks on the tissue slice) could be another solution to track optical tissue measurements \cite{Laughney2012ScatterAssessment, Unger2018MethodAssessment}. However, burn marks or other permanent markers can destroy the measured tissue and this can interfere with the following histopathology analysis, making this technique restricted to single points of interest as well. 

Using probe-fitting grids or molds is another way to track the optical tissue measurements locations without damaging the tissue \cite{DeBoer2019MethodDeformations}. But, the predefined grid locations can be insufficient since they will not always overlap with the aimed measurement location. Another method to localize measurement locations is video-tracking of an optical probe \cite{Horgan2021Image-guidedDelineation,Gkouzionis2022Real-timeSurgery}. Gorpas et al. proposed a live tracking technique for FLIm measurements by the incorporation of an aiming beam that allows localization during acquisition. A camera acquires the locations in a white light image, from which further optical analysis is feasible \cite{Marsden2020IntraoperativeLearning, Gorpas2016Real-TimeAccess}. The used wavelength range of this aiming beam is not affecting the FLIm acquisition. This technique is hard to incorporate for optical techniques where the probe needs to be in contact with tissue. Also, since this tracking method works with the use of an emitted blue light, broad-band spectroscopy such as DRS at certain wavelengths can be affected. Blocking the field of view of a camera can also result in failed tracking, which complicates the applicability of in vivo applications. In this paper, an improved method for tracking, registering, labeling and validating optical tissue measurements with histopathology is demonstrated. With the developed PPM system, it becomes possible to project any desirable number of measurement locations in a more controlled and automated manner without damaging or marking the specimen. This way human error are reduced, making this method more applicable compared to other tracking techniques available.

For this case study, lumpectomy specimens were processed in mega-cassettes to create a microscopic histology images of the complete tissue slices. It would be desirable to apply this framework, not only on lumpectomy specimen, but also within other oncologic domains in which optical tissue sensing technologies are investigated frequently and precision in correlation with histopathology is of great importance. However, when applying this framework to different types of tissue specimens, for example, colon or prostate, most often tissue slices must be subdivided in multiple cases since the tissue specimen is too big to process in a single case or hospitals have restrictions in adjusting standard histopathology processing protocols. In that case, microscopic histology images need to be reattached before using this framework, which can be complicated by tissue deformations. Before using this framework under those conditions, small adjustments to the methodology need to be taken into consideration to process tissue specimens and apply this framework in the most suitable way.  
The projection of POIs by the PPM system, due to base-plan and projector calibration, achieved high precision. However, the extraction of accurate tissue labels is dependent on the performance of the complete framework and relies also on the amount of tissue deformation that occurs during the histopathology processing of tissue slices. The developed automatic deformable registration is able to accurately register borders and inside structures when registering snapshot specimen images to histology images. However, when the tissue is deformed to a certain degree, the registration and following extraction of tissue labels will be affected. Tears, loss of tissue and holes making it difficult for the model to identify identical features to precisely overlay the images. This drawback is based on processes which are not related to this proposed framework, but do have an effect on the performance and need to be taken into consideration when using the obtained tissue labels for further development of tissue classification algorithms.

The performance of the automatic deformable registration is evaluated with the use of MI and DICE score, which determine differences in intensity level and the overlap between input images. We concluded that these matrices were the most suitable to determine registration accuracy between images in which it is difficult to find corresponding landmarks. However, other metrics such as target registration error can be explored to draw a more definite conclusion about the performance of the model. 

We would like to address that for the validation of optical tissue sensing techniques, and their further applicability in diagnostic tools, it is of great importance to correctly label the optically measured tissue with a ground truth. By using the proposed framework, manual and time-consuming tasks will be eliminated, which results in faster development of more robust and accurate classification algorithms.

\section{Conclusion}
\label{conclusion}
The developed Point Projection Mapping (PPM) system achieves accurate tracking of point-based optical tissue measurements performed on tissue specimens, making it widely applicable for the validation of optical tissue sensing technologies available. The proposed framework for the registration, validation and labeling of the tracked measurements with histopathology succeeded with high precision. 


\section*{Funding} The authors gratefully acknowledge the financial support of this research by the Dutch Cancer Society (Grant No. KWF 13443).



\section*{Acknowledgments} The authors would like to thank all surgeons and nurses from the Department of Surgery, all  pathologist assistants from the Department of Pathology for their assistance in processing specimens, the NKI-AVL core Facility Molecular Pathology \& Biobanking (CFMPB) for supplying NKI-AVL biobank material and all students that participated in this research for their time and effort. Research at the Netherlands Cancer Institute is supported by institutional grants of the Dutch Cancer Society and of the Dutch Ministry of Health, Welfare and Sport.


\bibliographystyle{unsrt}
\bibliography{references.bib}

\end{document}